\documentclass[useAMS]{mn2e}
\usepackage{graphicx}

\title[Eclipses and dust formation by WC9 stars]
{Eclipses and dust formation by WC9 type Wolf--Rayet stars}
\author[P. M. Williams]
       {P. M. Williams\thanks{Email: pmw@roe.ac.uk} \\
       Institute for Astronomy, University of Edinburgh, Royal Observatory, 
                 Edinburgh EH9 3HJ\\
   }

\date{Accepted 2014 August 27
      Received 2014 August 27;
      in original form 2014 August 12}

\pagerange{\pageref{firstpage}--\pageref{lastpage}}
\pubyear{2014}

\begin{document}

\maketitle

\label{firstpage}

\begin{abstract}
Visual photometry of 16 WC8--9 dust-making Wolf--Rayet (WR) stars during 
2001--2009 was extracted from the All Sky Automated Survey All Star Catalogue 
(ASAS-3) to search for eclipses attributable to extinction by dust formed in 
clumps in our line of sight. Data for a comparable number of dust-free WC6--9 
stars were also examined to help characterise the dataset. 
Frequent eclipses were observed from WR\,104, and several from WR\,106, 
extending the 1994--2001 studies by Kato et al. (2002a,b), but not supporting 
their phasing the variations in WR\,104 with its `pinwheel' rotation period. 
Only four other stars showed eclipses, WR\,50 (one of the dust-free stars), 
WR\,69, WR\,95 and WR\,117, and there may have been an eclipse by WR\,121, which 
had shown two eclipses in the past. No dust eclipses were shown by the `historic' 
eclipsers WR\,103 and WR\,113. The atmospheric eclipses of the latter 
were observed but the suggestion by David-Uraz et al. that dust may be partly
responsible for these is not supported. Despite its frequent eclipses, there is 
no evidence in the infrared images of WR\,104 for dust made in its eclipses, 
demonstrating that any dust formed in this process is not a significant 
contributor to its circumstellar dust cloud and suggesting that the same 
applies to the other stars showing fewer eclipses.

\end{abstract}


\newpage

\section{Introduction}

One of the earliest results from infrared (IR) astronomy was the discovery 
of `excess' IR radiation by heated circumstellar dust from a variety of 
stars having emission-line spectra. Amongst these were four WC9 type 
Wolf--Rayet (WR) stars, Ve2--45 (= WR\,104), HD 313643 (= WR\,106), LS~15 
(= WR\,119) and AS~320 (= WR\,121), observed by Allen, Harvey \& Swings (1972) 
and, at longer wavelengths, by Gehrz \& Hackwell (1974). Cohen, Barlow \& 
Kuhi (1975) measured optical--IR energy distributions of WR stars and showed 
that they could be matched by either free-free or graphite dust emission.

The enduring interest of dust formation by some WC type WR stars is the 
great difficulty of forming dust in such hostile environments: close to the 
stars, the stellar radiation fields heat any dust to above its evaporation 
temperature but, at greater distances where the radiation field is sufficiently 
diluted for the grains to survive in radiative equilibrium, the density in an 
isotropic WC stellar wind is too low to allow homonuclear dust condensation. 
This was recognised by Hackwell, Gehrz \& Grasdalen (1979) in the case of 
the WC7+O5 type episodic dust-maker HD 193793 (= WR\,140) and discussed by 
Williams, van der Hucht \& Th\'e (1987) for the WC8--9 stars making dust 
persistently. 
High-density structures in the WC winds are required to allow dust formation. 

Two classes of high-density structure are suggested by the observed incidence 
and phenomenology of dust formation: large-scale structures generated by the 
compression of the WC wind where it collides with the wind of a luminous 
companion in a binary system -- a colliding wind binary (CWB) -- and isolated 
clumps in stellar winds probably related to those observed spectroscopically 
by Moffat et al. (1988), Moffat \& Robert (1991) and L\'epine \& Moffat (1999). 
Usov (1991) modelled CWBs and suggested that very high compression factors 
($10^{3} - 10^{4}$) could be produced in WR\,140 if the heated and compressed 
wind was able to cool efficiently. The IR light curves show that dust 
formation in WR\,140 occurs only very near periastron passage in its highly 
elliptical 8-y orbit (Williams et al. 1990), 
presumably related to the increase by factor of $\sim$ 40 of the pre-shock 
wind density at the wind-collision region (WCR) around this phase (Williams 1999). 
Such phase-locking occurs in other episodic dust-making systems such as the 
$P = 13$-yr WC7+O system HD 192641 (= WR\,137) (Williams et al. 2001, Lef\`evre 
et al. 2005).

The connection between dust formation by the WC8--9 stars and their movement in 
binary orbits comes from IR images of the dust which show rotating `pinwheel' 
structures around systems like WR\,104, interpreted to be dust formed and 
ejected in a stream to one side of a rotating binary system observed at a low
orbital inclination (Tuthill, Monnier \& Danchi 1999, Tuthill et al. 2008). 
Further dust `pinwheels' have been observed around other WC8--9 stars, 
(cf. Monnier et al. 2007) -- but binary orbits are required to confirm the picture.

Dust formation in clumps in WR winds deduced from their eclipsing the light of 
the WR star was first suggested independently from spectroscopy of WR\,104 by 
Crowther (1997) and from photometry of WR\,121 and WR\,103 by Veen et al. (1998).
Compared with earlier observations, the spectrum of WR\,104 observed in 1996 
showed selective obscuration of the high-ionization emission lines, formed 
in the core of the WC9 wind, allowing Crowther to set a stringent limit on 
the projected size ($\sim 20 R_*$) of the obscuring dust clump 
using a model of the WR\,104 wind. 
Spectra observed by Williams \& van der Hucht (2000, hereafter WH00) showed 
that the strengths of the C\,{\sc ii} and C\,{\sc iii} lines had partly recovered 
from Crowther's 1996 eclipse by mid-1997, and that they had been near `normal' in 
mid-1995, a year before it.

Tighter timing of a dust eclipse was provided by Veen et al. (1998), who 
observed WR\,121 during and emerging from an 0.8-mag eclipse in 1990. The 
flux took about 10~d. to recover to its normal level. 
Multi-colour observations showed that the eclipse was deeper at shorter 
wavelengths, suggesting that it was caused by extinction by dust particles. 
A shallower (0.4-mag.) eclipse of WR\,103 also showed colour changes 
indicative of varying dust extinction. This star had previously shown 
isolated eclipses (Massey, Lundstr\"om \& Stenholm 1984), but they were not 
ascribed to dust formation. Another deep eclipse by WR\,121 in {\em Hipparcos} 
$Hp$ data was observed by Veen et al. and Marchenko et al. (1998), but its 
time-scale was not well defined because the observations of this star did 
not cover the whole event. 

Better defined is the deep ($\sim 2.9$-mag.), apparently multiple, eclipse by  
WR\,106 observed by Kato et al. (2002a) in 2000 in a photometric study running 
from 1994 to mid-2002. The observations were in a single passband, like 
the $Hp$ data, so interpretation in terms of dust absorption is by analogue 
with the multi-wavelength observations of eclipses.
On a smaller scale, variable circumstellar extinction by dust was invoked by 
Fahed, Moffat \& Bonanos (2009) to explain the low amplitude (0.06-mag.) 
variations of WR\,76 observed in their two-band ($V$ and $I$) photometric 
survey of 20 WC8--9 stars. 

Kato et al. (2002b) observed significant (2.7-mag.) variations in the light 
from WR\,104, with the star at its faintest near the time of Crowther's 
spectroscopy, supporting the suggestion that the variations were caused by 
variable dust extinction. They found a period of 241~d. in their data, close 
to that of the pinwheel orbit, and suggested that the dust eclipses were 
phase-locked to the orbit and caused by some of the dust plume creating the 
pinwheel entering our line of sight.  
Further optical photometry of WR\,104 in the All-Sky Automated Survey 
(ASAS-3, Pojma\'nski 2002) allowed reinvestigation of this suggestion.   
A dataset combining the observations by Kato et al. with the then available 
ASAS-3 observations, duly shifted to allow for the different passbands, showed 
(Williams 2008) no phasing of the optical photometry with the pinwheel period.

The ASAS-3 is now complete, prompting revisiting the variations of WR\,104, and 
extension of the study to a search for eclipses by other dust-forming WC8--9 stars. 
The prime goal is examination of the incidence of eclipses -- most of which had 
previously been discovered serendipitously -- as a step to investigating the 
significance of this mode of dust formation amongst WC8--9 stars as a possible 
alternative to formation in CWBs, with particular relevance to the WC8--9 stars 
without luminous companions. 
The ASAS-3 database is well suited to this, covering the region of the Galactic 
Plane occupied by the dust-making WC8--9 stars and containing observations made over 
a long time span (2001--2009) which, allowing for the gaps between the stars' observing 
seasons, provide the equivalent of $\sim$ 5~yr continuous photometry of each. The 
limiting magnitude is almost 15 and stars as bright as $V = 8.5$ can be observed 
without saturation (Pojma\'nski 2002), allowing inclusion of the brighter WC8--9 stars.  
As already noted, the interpretation of abrupt falls in single-passband photometry 
as obscuration by dust clumps is by analogue with the multi-wavelength studies of 
similar objects; but the present paper will proceed using this assumption, and 
follow Veen et al. (1998) in calling the phenomena eclipses. Such eclipses 
can tell us about varying dust absorption along a narrow column towards the stars; 
in contrast, the IR emission gives the total amount of dust around the stars close 
enough to be heated by the stellar radiation and re-emit in the IR. In this paper, 
the designation of stars as dust-makers or dust-free is on the basis of their IR 
spectral energy distributions (SEDs).

\section{Observational data}        

\begin{table}
\centering
\caption{Stars observed, with number of A-quality observations, median $V$ 
magnitude, standard deviation and aperture (pixels, 1 pixel = 15 arc sec) 
used for the photometry.}
\begin{tabular}{rrlrrlc}
\hline
 WR & HD/name  & Spectrum  & Nobs & $<V>$ & $\sigma V$ & Ap \\ 
\hline
 13 & Ve6--15  & WC6       &  614 & 12.86 &  0.09 & 2       \\
 14 & 76536    & WC7+?     &  557 &  8.82 &  0.04 & 4       \\
 15 & 79573    & WC6       &  562 & 10.65 &  0.04 & 4       \\ 
 23 & 92809    & WC6       & 1011 &  9.07 &  0.04 & 5       \\
 42 & 97152    & WC7+O7V   &  656 &  8.09 &  0.04 & 6       \\
 45 & LSS 2423 & WC6       &  199 & 14.11 &  0.17 & 2       \\
 50 & Th2--84  & WC7+OB    &  556 & 11.89 &  0.07 & 2       \\
 53 & 117297   & WC8d      &  990 & 10.53 &  0.04 & 4 \\
 56 & LS 8     & WC7       &  483 & 13.52 &  0.13 & 2       \\
 57 & 119078   & WC8       &  812 &  9.69 &  0.05 & 5       \\ 
 59 & LSS 3164 & WC9d      &  533 & 11.48 &  0.06 & 3 \\
 60 & 121194   & WC8       &  182 & 11.98 &  0.09 & 2 \\ 
 64 & BS 3     & WC7       &  319 & 14.04 &  0.16 & 2       \\
 65 & Wra 1297 & WC9d+OB   &  445 & 13.28 &  0.12 & 3 \\ 
 68 & BS 4     & WC7       &  543 & 13.24 &  0.12 & 2                    \\
 69 & 136488   & WC9d+OB   &  936 &  9.25 &  0.06 & 4 \\
 70 & 137603   & WC9d+B0I  &  919 &  9.80 &  0.03 & 5 \\
 73 & NS 3     & WC9d      &  114 & 14.45 &  0.20 & 2 \\
 77 & He3-1239 & WC8+OB    &  484 & 12.54 &  0.09 & 2 \\ 
 80 & LSS 3871 & WC9d      &  119 & 14.42 &  0.22 & 2 \\ 
 81 & He3-1316 & WC9       &  527 & 12.06 &  0.06 & 2                    \\
 92 & 157451   & WC9       &  602 & 10.27 &  0.05 & 2      \\ 
 93 & 157504   & WC7+O7-9  &  797 & 10.60 &  0.04 & 4                    \\
 95 & He3-1434 & WC9d      &  918 & 12.96 &  0.11 & 3 \\ 
 96 & LSS 4265 & WC9d      &  257 & 13.78 &  0.15 & 2 \\ 
101 & DA 3     & WC8       &  927 & 13.76 &  0.14 & 2                    \\  
103 & 164270   & WC9d      &  852 &  8.78 &  0.03 & 6 \\ 
104 & Ve2--45  & WC9d+B0   &  617 & 13.28 &  0.36 & 2 \\ 
106 & 313643   & WC9d      &  925 & 11.86 &  0.12 & 3 \\
113 & 168206   & WC8d+O8-9 &  733 &  9.13 &  0.03 & 5 \\
117 & IC14--22 & WC9d      &  429 & 13.11 &  0.14 & 3 \\ 
119 & Th\'e 2  & WC9d      &  422 & 12.02 &  0.06 & 2 \\ 
121 & AS 320   & WC9d      &  381 & 12.02 &  0.06 & 3 \\
125 & IC14--36 & WC7+O9III &  287 & 12.80 &  0.09 & 2 \\ 
\hline
\end{tabular}
\medskip
\noindent WR numbers assigned by van der Hucht et al. (1981)
\label{stars}
\end{table}

\begin{figure}                                                  
\centering
\includegraphics[width=8cm]{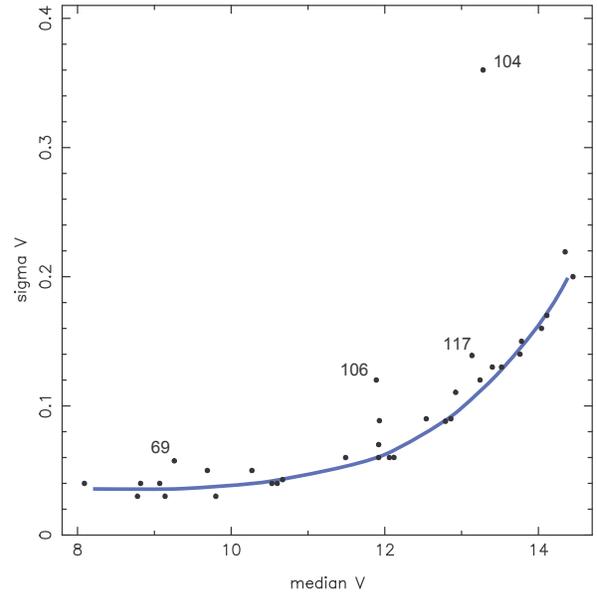}
\caption{Dispersion of ASAS-3 $V$ magnitudes ($\sigma V$) as a function 
of median $V$ of the stars in Table \ref{stars}. The line has the 
form of the equation for photometric error as a function of intensity 
derived by Schnurr et al. (2009), but fitted to the ASAS-3 data using 
the zero point, readout noise and flat-field noise as free parameters.}
\label{FVmedsig}
\end{figure}
 
Photometry for 16 dust-making WC8--9 stars was extracted from the ASAS-3 
All Star Catalogue\footnote{http://www.astrouw.edu.pl/asas/?page=aasc}. 
To avoid bias, the accessible dust-free WC8--9 stars (excluding WR\,88, 
which is a transitional WN/WC9 object, Williams, van der Hucht \& Rauw 2005) 
were added, together with a number of WC6--7 stars to help characterise 
the dataset. 

The ASAS-3 observations are flagged by quality (`A--D'), and only stars 
having at least 100 A-quality (best) data were used for this study. 
The photometry is provided through a range of apertures (2--6 pixels, 
where a pixel = 15 arc sec), with a suggested `default' aperture chosen 
on the basis of the object's brightness, smaller for the fainter objects. 
We began using the `default' apertures but in each case checked the 
scatter in the photomtery in other apertures; in a very few cases the 
scatter was found to less in an aperture other than the default, and 
the data measured through that aperture were used. 
The number of extracted observations, median $V$ magnitude, standard 
deviation and aperture size are given for each star in Table \ref{stars}.

The relation between the standard deviation and median $V$ magnitudes is 
shown in Fig\,\ref{FVmedsig},  which is consistent with the corresponding 
figure, based on very many more stars, in Pojma\'nski (2002). The photometric 
accuracy, 0.04--0.20 mag., is lower than that of some of the studies cited 
above, but entirely adequate for searching for $\Delta V \ga$ 0.5-mag. eclipses. 
In addition to observational errors, including calibration, the dispersions 
include the effects of any dust eclipses and also of the low-amplitude 
intrinsic variation observed in many WC9 stars (e.g. Moffat, Lamontagne \& 
Cerruti 1986; Balona, Egan \& Marang 1989; Fahed et al. 2009; van Genderen, 
Veijen \& van der Hucht 2013). These low-level variations are generally 
below the threshold of the ASAS-3 dataset and are not part of the present study.

\section{Results}        

\subsection{The dust-free WC stars}

\begin{figure}                                         
	
\centering
\includegraphics[width=8.5cm]{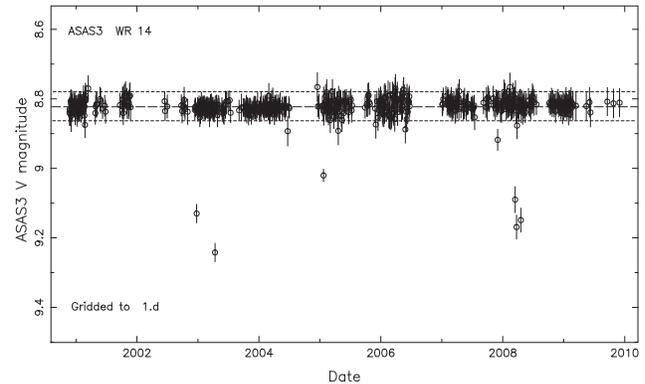}
\caption{Light curve of WR\,14 gridded to 1-day bins. 
Error bars on the individual points are the frame errors from the ASAS-3 
Catalogue, combined where necessary. The dashed line at $V = 8.82$ marks 
the median of the dataset and the dotted lines are spaced $\pm 1 \sigma$ 
about it (cf. Table \ref{stars}). The six points lying more than 3~$\sigma$ 
below the median are outliers: see text.}
\label{WR14all}
\end{figure}

\begin{figure}                                         
\centering
\includegraphics[width=8.5cm]{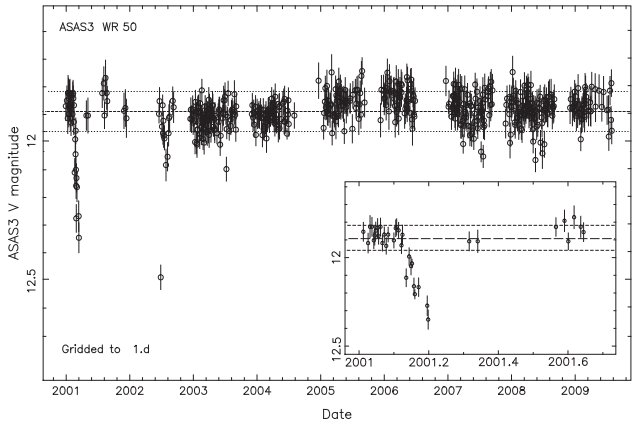}
\caption{As Fig.\ref{WR14all} for WR\,50 with appropriate median and errors. 
The 2001 fading during $\sim$ 30~d. is shown in detail in the inset; the 
faint magnitudes in 2002 and 2003 are isolated and considered outliers.}
\label{WR50comp}
\end{figure}

With the exception of WR\,125, these are stars which have never shown 
IR SEDs characteristic of dust formation. The episodic dust-maker 
WR\,125 (Williams et al. 1994) was included in this group because IR 
photometry in the 2MASS (Skrutskie et al. 2006), {\em Spitzer} GLIMPSE 
(Churchwell et al. 2003) and {\em WISE} (Wright et al. 2010) surveys, 
along with unpublished $L^{\prime}$ observations in 2006 and 2007, 
showed no recurrence of the 1990--93 dust-formation episode during the 
time span of the present study. It is possible that other stars in 
this group are unrecognised episodic dust makers observed during 
quiescence, since only a few WC8--9 stars have long IR photometric 
histories.                   

The light curve for each star was examined. To avoid clutter, the photometry 
was gridded to 1-d. intervals. Besides the dispersion ($\sigma$), 
which increases with faintness (cf. Fig.\,\ref{FVmedsig}), the light curves 
show a small number (about one per cent of the observations) of isolated 
points having $V \sim 0.5$ mag fainter than the median $V$. They are more 
conspicuous in the light curves of the brighter stars having less scatter; 
an example is shown in Fig.\,\ref{WR14all}. 
The light curves near each of the outliers were examined in detail to see 
if the outliers could represent real eclipses. Where the low values were not 
supported by the preceeding or following observations, they were  considered 
to be artefacts of the observations and unreliable -- apparently a 
characteristic of the dataset. 

With the single exception of WR\,50, none of the dust-free stars in 
Table \ref{stars} showed evidence for eclipse-like variation. The light 
curve of WR\,50 (Fig.\,\ref{WR50comp}) shows a well defined fading of 0.5~mag. 
over $\sim$ 30 d. in 2001. This was followed by a gap in the observing 
for $\sim$ 40~d., after which the flux had returned to its previous value. 
Short time scale, low amplitude photometric variation in 1988 was observed  
by van Genderen, Larsen \& van der Hucht (1990) and nightly fading in 1989 
reported by van Genderen et al. (1991); but reinvestigation of the latter by 
Veen \& Wieringa (2000) found that most, if not all, of this variation was 
not real. From IR observations in the 2MASS, GLIMPSE and {\em WISE} surveys, 
there is no evidence for dust emission in 2001--10 -- but this does not rule 
out the possibility that it is an episodic dust-maker observed during a long 
period of quiescence, like WR\,125. 
The spectrum of WR\,50 shows absorption lines (van der Hucht et al. 1981; 
Smith, Shara \& Moffat 1990) so WR\,50 is most probably a binary -- but 
apparently lacks an orbit.

\subsection{More eclipses in WR\,106}   

\begin{figure}                                                  
\centering
\includegraphics[width=8.5cm]{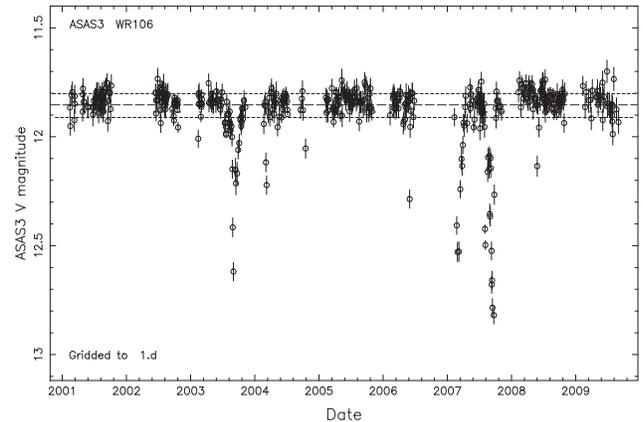}
\caption{As Fig.\ref{WR14all} for WR\,106; in this case the $\sigma$ for 
the dotted lines (0.06 mag.) was that read from the dispersion curve 
(Fig.\,\ref{FVmedsig}) for the brightness of WR\,106.}
\label{WR106all}
\end{figure}

\begin{figure}                                                  
\centering
\includegraphics[width=8.5cm]{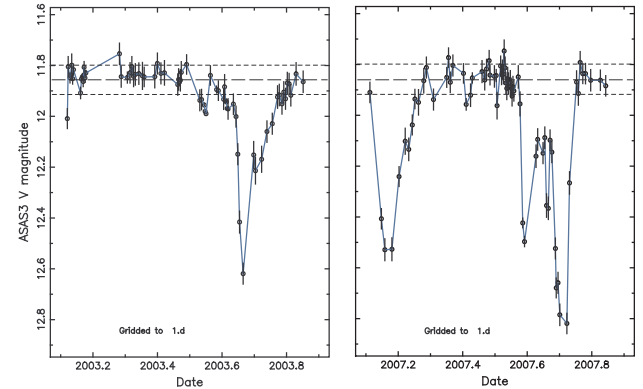}
\caption{Enlarged view of the 2003 and 2007 eclipses by WR\,106. The faint 
lines (blue in the on-line version) through the points are drawn to guide 
the eye and are not a model.}
\label{EC106}
\end{figure}

The ASAS-3 light curve (Fig.\,\ref{WR106all}) of WR\,106 shows deep eclipses 
in 2003 and 2007, the second eclipse in 2007 having a double minimum 
separated by $\sim$ 46 d. with perhaps a third, lower amplitude, one between 
them (Fig.\,\ref{EC106}). This invites comparison with the complex eclipse 
(in sparser data) observed in 2000 by Kato et al. (2002a), having at least two 
components, with minima separated by $\sim$ 56 days, but a larger amplitude 
($\sim 2.9$~mag), than that of the 2007.7 minimum ($\Delta V \simeq 1$ mag.). 

Kato et al. (2002a) also reported the marginal detection of a shallower 
($\sim$ 0.6-mag.) eclipse in 2002 May; unfortunately, the ASAS-3 observations 
in 2002 began in mid-June so this event cannot be compared in the two datasets. 
A possible low-amplitude ($\sim$ 0.4-mag.) eclipse in the ASAS-3 data could 
be the two fainter magnitudes in 2004.2 which were observed within 4~d. of 
each other and are therefore probably not observational outliers; the 
flux was near normal 2~d. before the first observation and again 4~d. 
after the second, making it a very brief event. The other low points 
in 2006 and 2008 in the light curve are isolated outliers.

The Kato et al. (2002a) and ASAS-3 datasets overlap in 2001--2, so that the 
two taken together provide coverage from 1994.1 to 2009.8, but with a 
number of gaps, mostly seasonal. To estimate the completeness of coverage 
with respect to discovery of deep eclipses like those observed, we note 
that the 2003 eclipse and the first in 2007 each lasted about 60~d., 
so that at least part of such an eclipse would still be observed if there 
was a gap of no more than 0.1~yr. in the observations. Omitting the larger 
gaps in coverage, we then have the equivalent of 9.9~yr of 
continous photometric monitoring during which six deep minima, four of 
them in close pairs, were observed. If the six minima were independent 
events, their distribution is far from Poissonian. 
For example, if we divided the 9.9 period into 33 intervals of 0.3~yr, an 
interval long enough to include the double eclipses but not so long as to 
include both the single and double eclipses in 2007, we have a rate of 
6/33 = 0.182 eclipses per interval. A Poissonian distribution would 
have 27.5 intervals with no eclipse, 5.0 with one eclipse, 0.45 with two 
eclipse, etc. -- far from what is observed. 
The numbers are too small for a conclusion, but this suggests that the 
double eclipses are integral events rather than the close proximity of 
independent single eclipses. This suggestion is supported by the apparent 
minor eclipse (depending on two observations separated by 2~d.) between 
the two components of the 2007 double eclipse, and would be strengthened 
if the apparent eclipses on the rising branches of the 2003.7 and 2007.2 
eclipses were real (Fig.\,\ref{EC106}).

The 2003 and the first 2007 eclipses show a slower recovery than fading, 
36 and 44~d., compared with fadings of 0.6 mag in 17 and 10 d. respectively.
The recovery from the double eclipse was much faster, 17~d. to rise by 0.9~mag.
If we interpret the rise times as the lifetimes of the dust clumps as they 
rise through the stellar wind in our direction, the simplest explanation of 
the shorter lifetimes is that the trajectories of the clumps make a small 
angle to our line of sight and they pass out of it sooner than if the 
trajectories were exactly aligned to our line of sight. 

The particular interest of formation of dust in clumps by WR\,106 is that it 
is not obviously a CWB: WH00 searched for, but did not observe, Balmer 
absorption lines attributable to an OB companion to the WC9 star in its spectrum. 
If WR\,106 is not a CWB, its circumstellar dust must be formed in some other process.

\subsection{Frequent eclipses in WR\,104}       

\begin{figure}                                                  
\centering
\includegraphics[width=8.5cm]{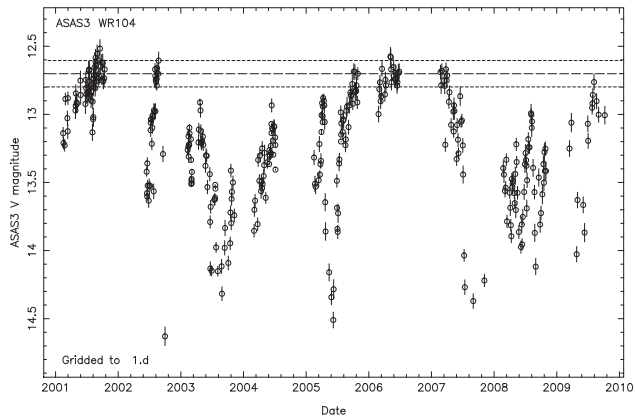}
\caption{Light curve of WR\,104, data gridded to 1-day bins. The dashed 
line at $V = 12.70$ is {\em not} the median, but an estimate of the uneclipsed 
flux from the light curve; and the $\sigma = 0.09$ used for the dotted lines 
spaced $\pm 1 \sigma$ about it is read from the dispersion curve for $V = 12.7$. }
\label{WR104all}
\end{figure}

\begin{figure}                                                  
\centering
\includegraphics[width=8cm]{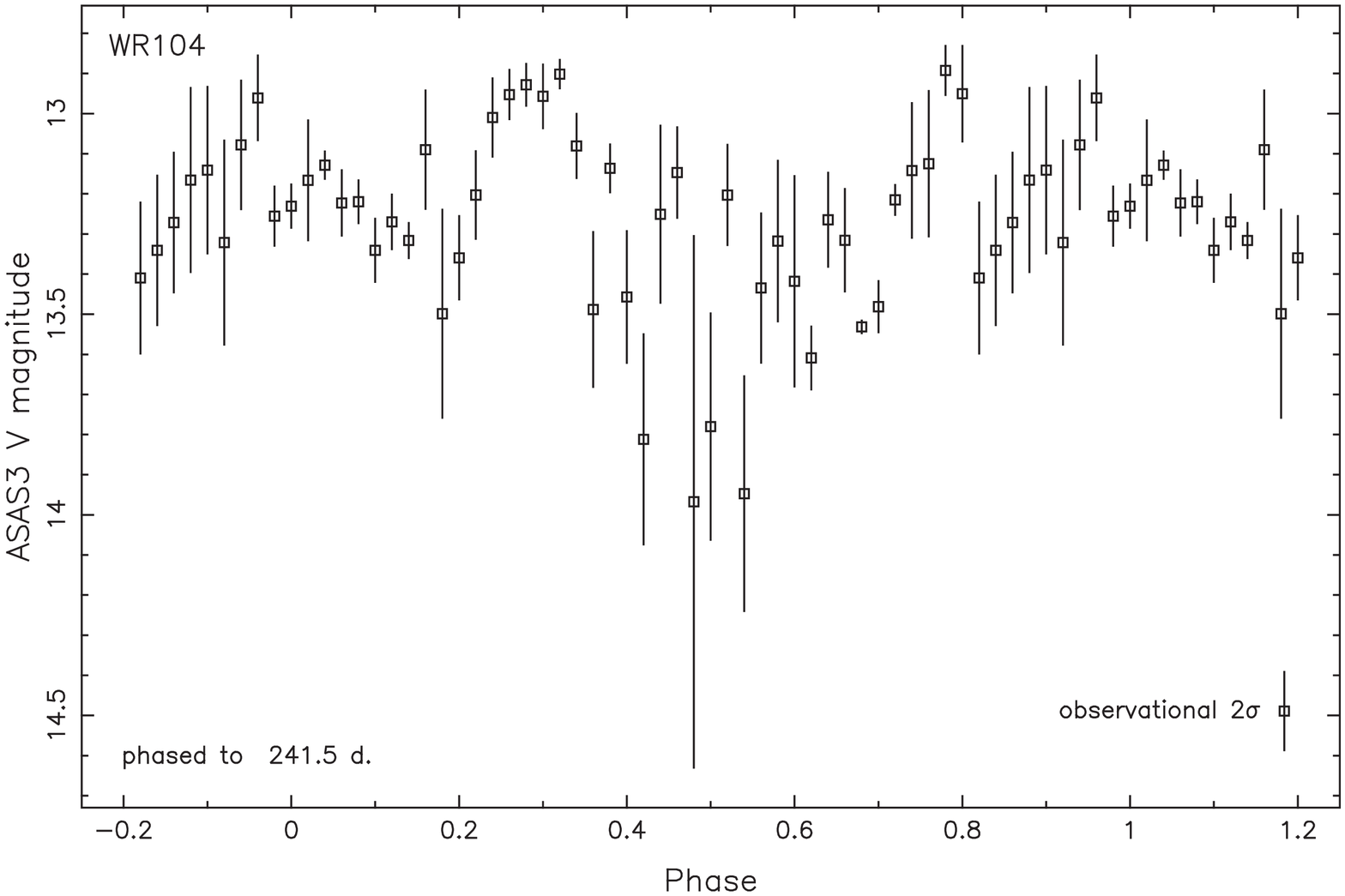}
\caption{Photometry of WR\,104 phased to the 241.5-d pinwheel period in 
50 0.02-P bins with zero phase at the first observation (JD 245\,1949.88). 
The error bars are $\pm 1 \sigma_{\rm m}$, and the expected observational 
error read from the dispersion curve is marked in the corner.} 
\label{Ph104}
\end{figure}

\begin{figure}                                                  
\centering
\includegraphics[width=8cm]{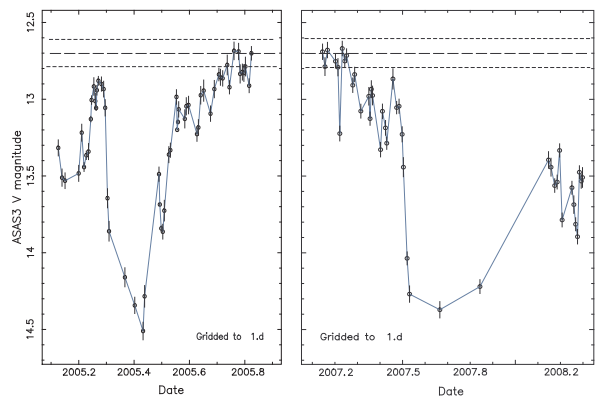}
\caption{Variation of WR\,104 in 2005 and 2007--8, data gridded to 1-day 
bins, showing eclipse in 2005 and a more prolonged minimum in 2007. 
Unfortunately, the data in the second half of 2007 are very sparse and 
there is a 110-d gap between the last 2007 observation and the first 
in 2008. As in Fig.\,\ref{EC106}, the faint lines through the points are 
drawn only to guide the eye.}
\label{EC104}
\end{figure}

The ASAS-3 light curve of WR\,104 (Fig.\,\ref{WR104all}) is reminiscent of 
that observed by Kato et al. (2002b) during 1994--2001 in showing almost 
continuous variation, and it differs significantly from all the other stars 
observed in not having long runs of apparently constant flux. 
Evidently, WR\,104 suffers more eclipses than any other WC9 star observed in this 
or previous studies, apparently an almost continuous string of eclipses having 
a range of depths, and sometimes overlapping. 
Another possible explanation for the variations is variable extinction caused 
by some of the dust formed in the colliding-wind system entering our line of 
sight, as proposed by Kato et al. (2002b). They found a period equal to the 
`pinwheel' rotation period in their photometric data but, as noted above, 
this was not supported in a preliminary study using their photometry and 
some early ASAS-3 data (Williams 2008).  
It is now possible to repeat the period search using the whole ASAS-3 
dataset for WR\,104, 617 observations compared with 146 used by Kato et al. 
The datasets were not combined this time as the Kato et al. data are of 
lower accuracy (0.2--0.3 mag.) than the ASAS-3 data, and are in a different 
photometric band (unfiltered 400--600 nm). 
A search for periods in the range 200--300 d. using phase dispersion 
minimization (PDM, Stellingwerf 1978) showed no evidence for a period near 
that (241.5~d., Tuthill et al. 2008) of the `pinwheel' rotation. 
The data were then phase-folded with the 241.5-d. period (Fig.\,\ref{Ph104}), 
giving the PDM statistic $\Theta = 0.75$, and do not show the large-scale 
phased variation of extinction found by Kato et al. The latter used 15 phase 
bins for their phased light curve in contrast to the 50 used here, which 
seems the minimum number needed to capture the variations of WR\,104. 
Inspection of the light curve shows well-defined changes of 0.1--0.2 mag., 
compared with 0.1 mag. for the expected observational error, on the time scale 
of the P/50 phase bins, so wider phase bins would smother these variations.
Phased light curves using narrower binning were also investigated, but none 
showed systematic variation. The large dispersions seen in some of the phase 
bins is a consequence of stochastic occurence of deep eclipses independent 
of the orbital period.

There are so many eclipses that the median $V$ is not a good measure of the 
undisturbed flux level, which is estimated to lie near $V = 12.7$ from 
inspection of the light curve. 
Omitting gaps in the observing exceeding 0.1~yr, we have the equivalent of 
4.8~yr of continuous photometric coverage, during which WR\,104 was brighter 
than $V = 12.8$ (1 $\sigma$ below the estimated undisturbed level) for only 
about 0.64~yr, implying that eclipses were occurring for seven-eighths of the 
time. From the light curves, 14 eclipses were identified; some completely 
observed, and others inferred from well-defined but isolated brightening or 
fading at the ends of observing runs, of which three had amplitudes in the range 
1.5--2.0 mag., eight in 1.0--1.4 mag., and three (probably an underestimate) 
in 0.5--0.9 mag., giving an average rate of 3.4~yr$^{-1}$, also probably an 
underestimate.

The light curves show much wider minima than the eclipses shown by WR\,106, 
but this might be caused by the partial overlapping of individual events. 
For example (Fig.\,\ref{EC104}), there was a reasonably well defined eclipse 
in 2005, but its recovery appears to have been interrupted by another eclipse, 
so estimation of the recovery time and clump lifetime comes from measuring the 
brightening rates, giving $\sim$ 40~d.  
In 2007, there was an even broader minimum, but it is not obvious whether this 
was caused by a single clump or several -- the data in the second half of 2007 
are very sparse, with a gap of 65~d. between the last two observations, and one 
of 110-d. between the last and the first observations in 2008. During 2008, 
the light curve looks like noise, but closer examination shows $\sim$ 1-mag 
eclipses in 2008.4 and 2008.6. Again it was not possible to measure recovery 
times or clump lifetimes as the recoveries were interupted. Gaps in the 
coverage make other events harder to characterise, but the steady rise of 
0.9 mag. in 0.18~yr in 2002, recovery from an earlier eclipse, suggests a clump 
lifetime of at least 66~d.

\subsection {0bservations of other dust-making WC8--9 stars}

\begin{figure}                                                  
\centering
\includegraphics[width=8.5cm]{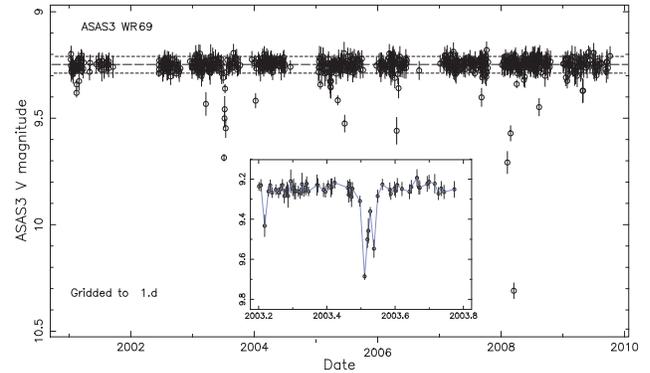}
\caption{As Fig.\ref{WR14all} for WR\,69 with appropriate median and errors.
The 2003.5 eclipse is shown in the inset; the other low points are all 
isolated outliers.}
\label{WR69comp}
\end{figure}

The photometric dispersion in WR\,69 seems anomalously high for its brightness 
(cf. Fig.\,\ref{FVmedsig}), and the light curve (Fig.\,\ref{WR69comp}) shows 
rather more outliers than the others. The dip in mid-2003 comes from seven 
observations (some gridded) and is probably a real, low-amplitude, eclipse. 
Inspection of all the other outliers showed that they were isolated observations. 
Low-amplitude ($\sim 0.02$-mag.) variability has been observed by Balona et al. (1989), 
and was found in {\em Hipparcos} $Hp$ data by Marchenko et al. (1998) and 
Koen \& Eyer (2002), with different periods, but phasing the ASAS-3 photometry 
to both of these periods did not reveal any systematic variation. 
Spectroscopy showed the WC9 star to have an early O-type companion (WH00), 
and a subsequent observation showed the lines in the two stars to have 
moved relative to each other, confirming WR\,69 as a spectroscopic binary 
(Williams et al. 2005).

\begin{figure}                                                  
\centering
\includegraphics[width=8.5cm]{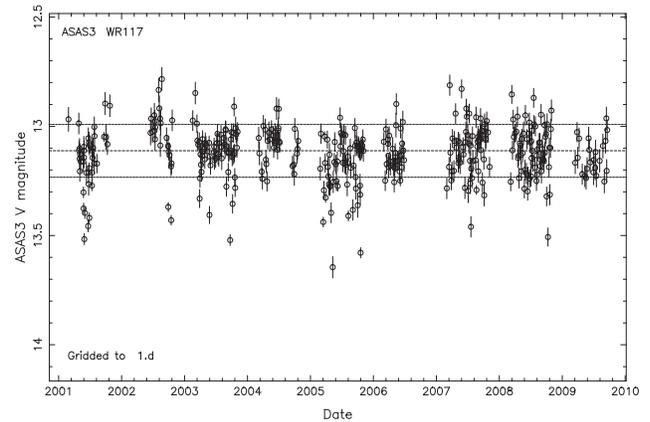}
\caption{As Fig.\ref{WR14all} for WR\,117 with appropriate median and errors.}
\label{WR117all}
\end{figure}

\begin{figure}                                                  
\centering
\includegraphics[width=8.5cm]{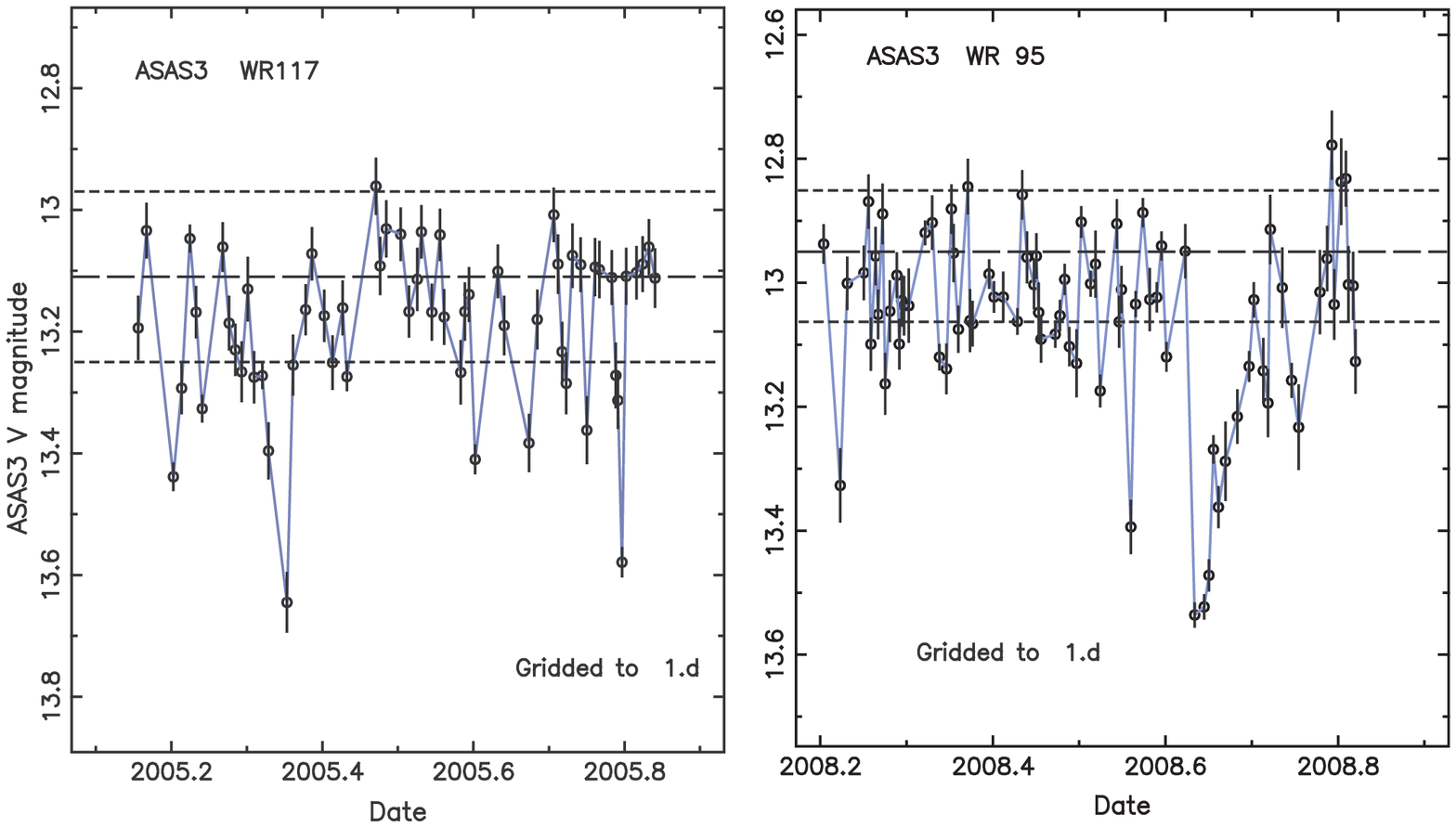}
\caption{Details of light curves showing eclipses of WR\,117 in 2005 (left) 
and WR\,95 in 2008 (right). Several isolated low points not considered to 
be eclipses are also visible. As in Fig.\,\ref{EC106}, the faint lines through 
the points are drawn only to guide the eye.}
\label{EC11795}
\end{figure}

\begin{figure}                                                  
\centering
\includegraphics[width=8.5cm]{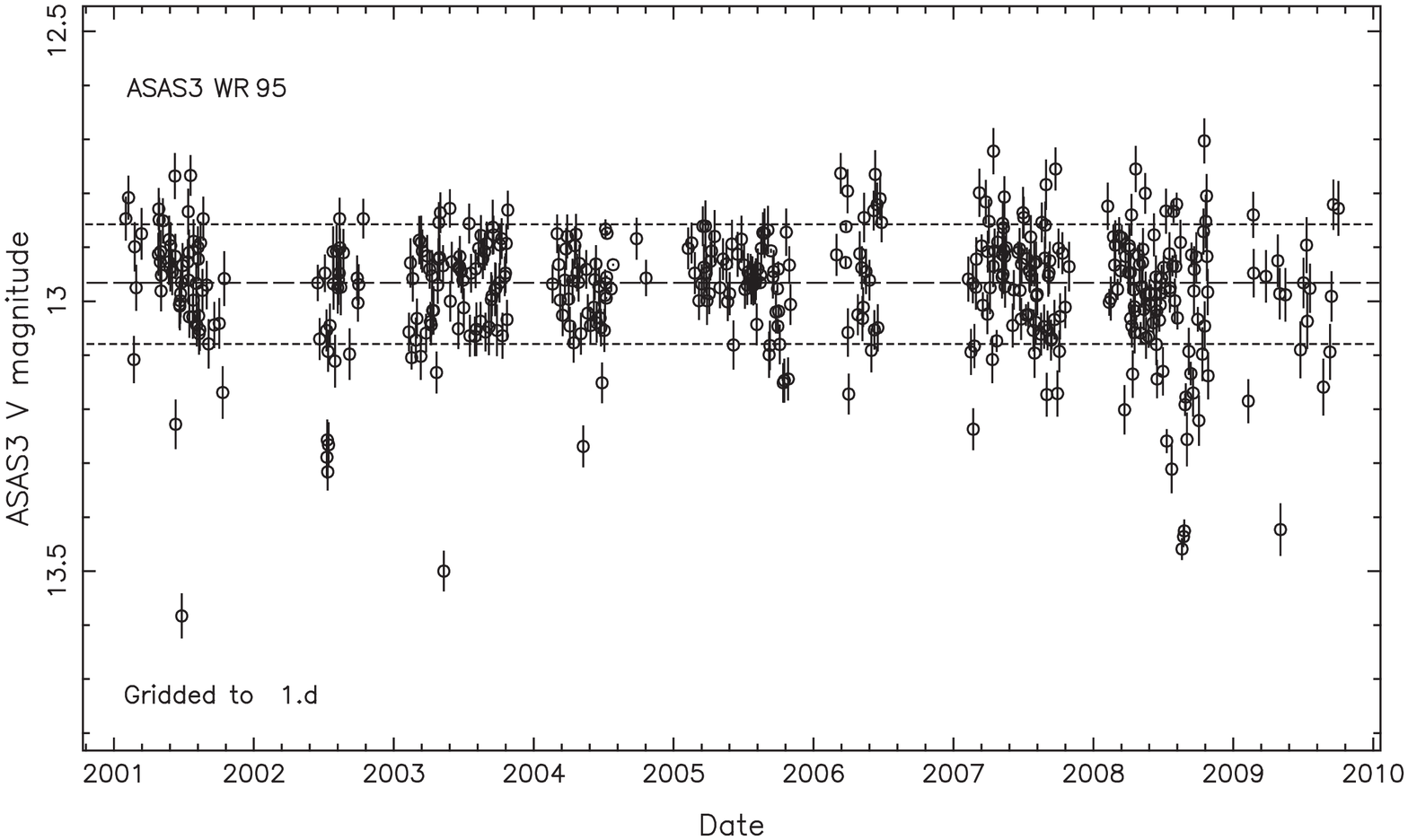}
\caption{As Fig.\ref{WR14all} for WR\,95 with appropriate median and errors.}
\label{WR95all}
\end{figure}

Another star showing slightly high dispersion for its magnitude is WR\,117, 
whose light curve (Fig.\,\ref{WR117all}) shows a fair amount of scatter and 
a possible 0.5-mag. eclipse defined by four points in 2005.4. This is shown 
at higher scale in Fig.\,\ref{EC11795}.

The light curve of WR\,95 (Fig.\,\ref{WR95all}) shows a 0.6-mag. eclipse  
in 2008.6 (Fig\,\ref{EC11795}), with a slow recovery ($\sim$ 40~d.); and a 
possible weak (0.3-mag.) eclipse in 2002.3. 

No dust eclipses were identified in the light curves of WR\,103 or WR\,113, 
which had shown eclipses in the past (see above), based on the equivalents 
of 3.8 and 5.2 yr of continuous monitoring. Phasing the photometry of 
WR\,113 (CV~Ser), omitting the 7 out of 733 magnitudes identified as 
isolated outliers, to the elements of David-Uraz et al. (2012) clearly 
showed a broad ($\sim 0.15$~P) 0.02-mag. atmospheric eclipse at phase 0.5, 
matching their observations and giving a measure of the quality of the 
ASAS-3 photometry of a bright star. 
David-Uraz et al. suggested that scattering by dust in the wind of the 
WC8 component could be a significant contributor to the depth of the eclipse, 
but this idea suffers from the problem that any dust would also absorb stellar 
radiation which would heat it -- to evaporation, if too close to the star. 
From the $a \sin i = 49 R_{\odot}$ measured by Niemela et al. (1996) and 
$\sin i = 0.979$ given by David-Uraz et al., the line of sight to the O~star 
can be seen to pass within $10 R_{\odot}$ of the WC8 star.    
This is very much closer to the star than the minimum distance at which carbon 
grains can survive the radiative heating, the inner edge of the WR\,113 dust 
cloud ($421 R_*$, where $R_*$ is the radius of a single star having the same 
luminosity as the WC8+O binary) modelled by Williams et al. (1987), or the 
corresponding distances, 20--30~au, observed and modelled for the WC9+B0 
system WR\,104 by Tuthill et al. (2008) and Harries et al. (2004), so it is 
most unlikely that dust can play a part in the atmospheric eclipses.

No well defined eclipse was shown by the third star to have shown eclipses 
in the past, WR\,121, during the equivalent of 3.8 yr of continuous monitoring. 
Two isolated points 0.2 mag. below median observed on successive nights, 
supported by a B-quality observation four nights earlier, at the end of 
the 2008 observing season might point to an eclipse but the coverage was 
too sparse to be certain. 

None of the other eight dust-makers showed evidence for an eclipse in their 
light curves or, indeed, any significant variation.

\section{Discussion}              

\subsection{The incidence of eclipses}

Apart from WR\,104 and WR\,106, in which at least 14 and 6 eclipses with 
amplitudes greater than 0.5~mag. were observed in the equivalents of 4.8 
and 9.9 yr of continuous monitoring respectively, the other 14 dust-making 
WC8--9 stars accounted for about four eclipses of lower amplitude in the 
equivalent of $\sim 65$ yr of continuous monitoring. This very great 
discrepancy in observed eclipse rates suggests a significant difference 
between the winds of WR\,104 and WR\,106 and those of the other WC8--9 
stars observed: greater than the high end of a continuous distribution 
of properties.

An alternative to the very uneven distribution of clump formation in WC9 winds 
is that all the stars form clumps at rates comparable to WR\,104 and WR\,106, 
but that the clumps are confined to narrowly collimated streams, and that only 
these two stars are favourably inclined to the observer. In this picture, the 
streams of clumps from the stars showing occasional eclipses have trajectories 
making slightly larger angles with our line of sight. The problem with this 
explanation is that it runs contrary to the spectroscopic study of the radial 
velocities of clumps in WR winds, which show their trajectories to have a 
range of angles to the line of sight (Moffat \& Robert 1991; Robert 1994; 
L\'epine \& Moffat 1999), and that it is most unlikely that the two sorts of 
clump would be formed in such different ways.

The eclipse rates can be compared with the circumstellar dust clouds, as 
measured by the fractions of the stellar ultraviolet--visible flux which 
they absorb and re-radiate in the IR. For WR\,104 and WR\,106, these 
are about 60 and 23 per cent respectively, compared with 1--10 per cent 
(Williams et al. 1987) for the other dust makers in this study. 
Amongst the latter, WR\,80 and WR\,96 re-radiate approximately 9.5 per cent 
of their stellar flux in the IR but have not shown eclipses, whereas the 
stars showing a few eclipses, WR\,69, WR\,95, WR\,117 and perhaps WR\,121 
have re-radiation rates between 1.7 and 5.8 per cent, and WR\,50 is not a 
known dust maker; so the correlation is not strong. 
Unfortunately, the other stars shown by Williams et al. (1987) to have 
high IR re-radiation rates, WR\,48a, WR\,76 and WR\,118, are all too faint 
for the present study (Fahed et al. (2009) did not observe any eclipses by 
WR\,76 in their study, but the photometry covered only 18~d.)
Fainter WC8--9 dust makers, including WR\,48a and WR\,76, should be 
observable in the currently progressing surveys such as 
ASAS-4\footnote{http://www.astrouw.edu.pl/asas/?page=asas4} and OGLE-IV 
(cf. Pietrucowicz et al. 2013). The photometric coverage of at least some 
of the stars in the present study should be extended to improve our knowledge 
of the incidence of eclipses in stars like WR\,121 and to investigate the 
possible double or multiple eclipses apparently shown by WR\,106.

To study the connection between the dust formed in clumps and that making up 
the circumstellar dust reservoirs responsible for the IR emission, consider 
what happens to the dust. The clump apparently starts close to the WC star 
(Crowther 1997, Veen et al. 1998) and moves out with the wind. The dust 
condenses quickly, judging by the fading times of the eclipses.
As the dust absorbs stellar radiation, it heats up and re-radiates in the IR 
until either it gets so hot that the dust evaporates, or the clump has moved 
away from the star sufficiently quickly that the geometric dilution of the 
stellar radiation heating it is no longer enough to heat it above evaporation 
temperature. By the end of the eclipse, either the dust has been destroyed, 
or the dust has survived but the clump has dissipated sufficiently that it 
no longer provides enough visual absorption to produce a measurable eclipse 
-- but, of course, continues absorbing and re-radiating in the IR.
There are two time-scales: the kinematic time for the dust to reach a safe 
distance (e.g. about 25 au in the case of WR\,104 from the modelling by 
Harries et al. 2004) which would be $\sim 35$~d. at an average speed of 
1200 km~s$^{-1}$, the terminal velocity of the wind. 
This may be optimistic, the initial acceleration of clumps is rather slow 
(e.g. L\'epine \& Moffat 1999), but radiation pressure will accelerate the dust 
to move faster than the wind in which it condensed. 
The other time-scale is that for the dust being heated in the stellar radiation 
field to reach its evaporation temperature. This needs to be modelled, particularly 
the properties of the original clump and the shielding it must provide to the 
condensing and newly formed dust -- which is beyond the scope of the present study. 
We can see, however, that the kinematic time-scale is comparable to the recovery 
times of eclipses, so it seems possible that dust formed in clumps is surviving 
long enough to reach the `safe' distance and contribute to the circumstellar dust 
reservoir. Simultaneous optical and IR observations during a dust eclipse would 
be invaluable in measuring the brightening in the IR and temperature of the dust 
clump as it moves out in the stellar wind. One star, WR\,70 was observed frequently 
in the IR during the present study, varying with a range of $\Delta K \simeq 0.5$ 
(Williams et al. 2013), but its ASAS-3 $V$ magnitudes showed no eclipses or other 
variation, so the variations in dust emission must be ascribed to changes in its
WC9\,+\,B0I wind-collision system.

\subsection{Dust formation by WR\,104}

The absence of variable extinction associated with the orbital motion in WR\,104 
is unsurprising considering the geometry. We know from the spectroscopy in 
1995--97 that dust was in front of the WC9 star and not the OB companion. 
The winds collide between the stars, and the dust is formed further from 
the WC9 star, beyond the OB companion, in a well-collimated beam 
(opening angle $40\degr$, cf. Harries et al. 2004) in the orbital plane, 
which lies almost in the plane of the sky given the low inclination, 
$i \simeq 11\degr$. 
Only in an inclined system like WR\,140, is some of the CWB dust modelled to 
pass in front of the star (Williams et al. 2009) and cause the minor eclipses 
observed shortly after periastron passage (Marchenko et al. 2003), when the 
CWB dust formation was briefly active. The high eccenticity of its orbit 
results in the projected position angle of the wind-collision system swinging 
through more than $180\degr$ on the sky in only $0.02P$ around periastron and 
the dust to be spread very unevenly around the stars.

The IR images of WR\,104 (Tuthill et al. 2008) can be used to infer the  
relative importance to the circumstellar dust reservoir of formation in 
clumps and formation in the WC9+OB wind-collision system responsible for 
the dust pinwheel. 
If dust made in clumps responsible for the eclipses is significant, we would 
expect to see evidence in the form of extended, possibly clumpy, diffuse 
emission centred on the star in the IR images. The very observation of pinwheel 
structures, however, demonstrates that most of the dust emission is in the 
pinwheel; the composite 2-$\umu$m image (Tuthill et al. 2008, fig.\,8) 
shows no dust emission above 0.1 per cent of the peak between the arms. 
This demonstrates that dust formation in clumps is not an important contributor 
to the total dust being formed by WR\,104 -- either because comparatively 
little dust is being made in this process, or because the dust in the clumps 
does not survive long enough to reach the `safe' distance. 

If instead of isotropic clump formation, we consider that WR\,104 owes its high 
rate of observed eclipses to a collimated beam and our favoured viewing angle, 
we might expect the IR images to show a bright spot in the centre: emission by 
a column of surviving clump dust. There is no evidence for this; on the contrary, 
the images show a `standoff distance' of 13~mas (Tuthill et al. 2008), the `safe' 
distance above, between the brightest dust pixel and the spiral centre, 
interpreted as the stars, where no emission is seen. 

In either case, it is evident that the star having by far the highest rate of 
eclipses is not receiving a significant contribution to its circumstellar dust 
cloud by this process. We need to know more about the WR\,104 eclipses from 
multi-colour photometry over a wide wavelength range to make sure that the 
extinction really is by dust and to measure its properties. 
Large amplitude ($\Delta r = 0.87, \Delta i = 0.76$) variability was 
found in the first results from the Bochum southern Galactic Disk survey 
(Haas et al. 2012) and the combination of such photometry with contemporaneous 
$V$ in the ASAS-4 survey over the next few years would be valuable.
Also, WR\,104 would be a good candidate for simultaneous IR and optical photometry 
to determine the behaviour of the dust on the recovery branches of the eclipses.

\section{Conclusions}

Of the 16 dust-making WC8--9 stars surveyed, most of the eclipses were observed 
from just two stars, WR\,106 and WR\,104, while the other 14 stars showed only 
3--4 eclipses, none as deep as those in WR\,104 or WR\,106. It is evident from 
the IR images of WR\,104 that any dust formed in the eclipses is not a significant 
contributor to its circumstellar dust cloud. 
The implication of this is that dust formed in occasional eclipses by the other 
WC8--9 dust makers is also an insignificant contributor to their dust clouds, 
so that the question of how WC9 stars which are not members of CWBs make their 
dust still remains open. 

The two stars with the highest eclipse rates also have the most luminous dust 
clouds in the IR and, although dust formed in clumps may not contribute to the 
circumstellar dust directly, CWBs having higher rates of clumps in the WC9 winds 
might be more efficient dust makers.

\section*{Acknowledgements}

This work is based on the All Sky Automated Survey All Star Catalogue and 
it is a pleasure to thank Grzegorz Pojma\'nski for helpful correspondence. 
It has made use of the NASA/IPAC Infrared Science Archive, which is operated 
by the Jet Propulsion Laboratory, California Institute of Technology, under 
contract with the NASA, and the SIMBAD and VizieR databases, operated at the CDS, 
Strasbourg. I am grateful to the Institute for Astronomy and UK Astronomy 
Technology Centre for continued hospitality and access to the facilities of 
the Royal Observatory Edinburgh.


\begin{thebibliography}{99}

\bibitem[AHS]{}  Allen D. A., Harvey P. M., Swings J. P., 1972, A\&A, 20, 333

\bibitem[photWR]{} Balona L. A., Egan J., Marang F., 1989, MNRAS, 240, 103  

\bibitem[GLIMPSE]{} Churchwell E., et al., 2009, PASP, 121, 213  

\bibitem[CBK]{} Cohen, M., Barlow, M.J. \& Kuhi, L.V. 1975, A\&A, 40, 291

\bibitem[ec104]{} Crowther P. A., 1997, MNRAS, 290, L59  

\bibitem[WR113]{} David-Uraz A., et al., 2012, MNRAS, 426, 1720  

\bibitem[varWC9]{} Fahed R., Moffat A. F. J., Bonanos A. Z., 2009, MNRAS, 392, 376

\bibitem[GH74]{} Gehrz R. D., Hackwell J.A., 1974, ApJ, 149, 619

\bibitem[Bochum]{} Haas M., Hackstein M., Ramolla M., Drass H., Watermann R., Lemke R., 
                   Chini R., 2012, AN, 333, 706

\bibitem[HGG140]{} Hackwell J. A., Gehrz R. D., Grasdalen G. L., 1979, ApJ, 234, 133

\bibitem[3D104]{}  Harries T. J., Monnier J. D. Symington N. H. Kurosawa R., 2004,
                   MNRAS, 350, 565

\bibitem[Kato106]{} Kato T., Haseda K., Takamizawa K., Yamaoka H., 2002a, A\&A, 393, L69

\bibitem[Kato104]{} Kato T., Haseda K., Yamaoka H., Takamizawa K., 2002b, PASJ, 54, L51

\bibitem[HippVar]{} Koen C., Eyer L., 2002, MNRAS, 331, 45  

\bibitem[Orbit137]{} Lef\`evre L., et al. 2005, MNRAS, 360, 141

\bibitem[clumpsII]{} L\'epine S., Moffat A. F. J., 1999, ApJ, 514, 909

\bibitem[HipPhot]{} Marchenko S. V., et al., 1998, A\&A, 331, 1022  

\bibitem[WR140]{} Marchenko S. V., et al., 2003, ApJ, 596, 1295

\bibitem[hist103]{} Massey P., Lundstr\"om I., Stenholm B., 1984, PASP, 96, 118

\bibitem[WR103v]{} Moffat A. F. J., Lamontagne R., Cerruti M., 1986, PASP, 98, 1170

\bibitem[var]{} Moffat A. F. J., Robert C., 1991, in in van der Hucht K. A., Hidayat B., eds, 
                    IAU Symp. No. 143, Wolf-Rayet Stars and Interrelations with other Massive 
                    Stars in Galaxies, Kluwer Academic Publishers, Dordrecht, p.~109

\bibitem[Clumps1]{} Moffat A. F. J., Drissen L., Lamontagne R., Robert C., 1988, ApJ, 334, 1038  

\bibitem[sizes]{} Monnier J. D., Tuthill P. G., Danchi W. C., Murphy N., Harries T. J., 
                  2007, ApJ, 655, 1033

\bibitem[orb140]{} Monnier J. D., et al., 2011, ApJL, 742, L1   

\bibitem[CV Ser]{} Niemela V. S., Morrell N. I., Barba R. H., Bosch G. L., 1996, 
                   Rev. Mex. Serie de Conferencias, 5, 100

\bibitem[OGLE]{} Pietrukowicz P., et al., 2013, Acta Astr., 63, 379  
				  
\bibitem[ASAS-3]{} Pojma\'nski G., 2002, Acta Astr. 52, 397

\bibitem[blobs]{} Robert C., 1994, Ap\&SS, 221, 137

\bibitem[R136]{} Schnurr O., Chen\'e A.-N., Casoli J., Moffat A. F. J., St-Louis N., 
                 2009, MNRAS, 397, 2049

\bibitem[2MASS]{} Skrutskie M. F., et al., 2006, AJ, 131, 1163

\bibitem[SSM1]{} Smith L. F., Shara M., Moffat A. F. J., 1990, ApJ, 348, 471

\bibitem[PDM]{}   Stellingwerf R. F., 1978, ApJ, 224, 953

\bibitem[WR104]{} Tuthill P. G., Monnier J. D., Danchi, W. C. 1999, Nature, 398, 486

\bibitem[WR104b]{} Tuthill P. G., Monnier J. D., Lawrance N., Danchi W. C., Owocki S. P., 
                   Gayley K. G., 2008, ApJ, 675, 698

\bibitem[DustTh]{} Usov V. V. 1991, MNRAS, 252, 49

\bibitem[6thCat]{} van der Hucht K. A., Conti P. S., Lundstr\"om, I., Stenholm B., 
                   1981, Space Sci. Rev., 28, 227

\bibitem[WR50etc]{} van Genderen A. M., Larsen I., van der Hucht K. A., 1990, A\&A, 229, 123

\bibitem[varBali]{} van Genderen A. M., et al., 1991, in van der Hucht K. A., Hidayat B., eds, 
                    IAU Symp. No. 143, Wolf-Rayet Stars and Interrelations with other Massive 
                    Stars in Galaxies, Kluwer Academic Publishers, Dordrecht, p.~129

\bibitem[WR103etc]{} van Genderen A. M., Veijen S. R. G., van der Hucht K. A., 2013, Ap\&SS, 345, 133

\bibitem[radio50]{} Veen P. M., Wieringa M. H., 2000, A\&A, 363, 1026

\bibitem[Eclipse121]{} Veen P. M., van Genderen A. M., van der Hucht K. A., Li, A., 
                       Sterken C., Dominik C., 1998, A\&A, 329, 199

\bibitem[IAU193]{} Williams P. M.  1999, in:  van der Hucht K. A., Koenigsberger  G.,
                   Eenens P. R. J., eds, Proc. IAU Symp. No. 193, Wolf-Rayet Phenomena 
                   in Massive Stars and Starburst Galaxies, Astron. Soc. Pacific, 
                   San Francisco, p.~267

\bibitem[Virpi]{} Williams, P. M., 2008, Rev. Mex. Serie de Conferencias, 33, 71

\bibitem[SAAO]{} Williams P. M., van der Hucht K.A., 2000, MNRAS, 314, 23 (WH00) 

\bibitem[WHT]{} Williams P. M., van der Hucht K. A., Th\'e P. S., 1987, A\&A, 182, 91 

\bibitem[Paper1]{} Williams P. M., van der Hucht K. A., Pollock A. M. T., Florkowski D. R., 
                   van der Woerd H., Wamstecker W. M.,  1990, MNRAS, 243, 662

\bibitem[WR125b]{} Williams P. M., van der Hucht K. A., Kidger M. R., Geballe T. R., 
                   Bouchet P., 1994, MNRAS, 266, 247

\bibitem[WR137ii]{} Williams P. M., et al., 2001, MNRAS, 324, 156  

\bibitem[LiegeWC9]{} Williams P. M., van der Hucht K. A., Rauw G., 2005, in: Rauw G., 
                     Naz\'e Y., Blomme R., Gosset E., eds, Massive Stars and High-Energy 
                     Emission in OB Associations, a workshop of the JENAM 2005, 
                     `Distant Worlds', Li\`ege 2005, p.~65

\bibitem[WR140im]{} Williams P. M., et al., 2009, MNRAS 395, 1749

\bibitem[WR70]{} Williams P. M., van der Hucht K. A., van Wyk F., Marang F., Whitelock P. A., 
                 Bouchet P., Setia Gunawan D. Y. A., 2013, MNRAS, 429, 494

\bibitem[WISE]{}  Wright E. L., et al. 2010, AJ, 140, 1868

\end{thebibliography}
\end{document}